\documentclass[twocolumn,aps,prl,preprintnumbers]{revtex4-1}
\usepackage{bbm}
\usepackage[latin9]{inputenc}
\usepackage{amsmath}
\usepackage{amssymb}
\usepackage{graphicx}
\usepackage{mathrsfs}
\usepackage{amsfonts}
\usepackage{amsthm}
\usepackage{color}
\usepackage{txfonts}
\usepackage[colorlinks=true,citecolor=blue,linkcolor=blue,urlcolor=blue,anchorcolor=blue]{hyperref}%
\usepackage{pifont}
\hypersetup{colorlinks=true,citecolor=blue,linkcolor=blue,urlcolor=blue}
\providecommand{\U}[1]{\protect\rule{.1in}{.1in}}
\makeatletter
\@ifundefined{textcolor}{}
{
\definecolor{BLACK}{gray}{0}
\definecolor{WHITE}{gray}{1}
\definecolor{RED}{rgb}{1,0,0}
\definecolor{GREEN}{rgb}{0,1,0}
\definecolor{BLUE}{rgb}{0,0,1}
\definecolor{CYAN}{cmyk}{1,0,0,0}
\definecolor{MAGENTA}{cmyk}{0,1,0,0}
\definecolor{YELLOW}{cmyk}{0,0,1,0}
}

\makeatother
\begin{document}
\title{Simulating Thiele's Equation and Collective Skyrmion Dynamics in Circuit Networks}
\author{Huanhuan Yang}
\author{Lingling Song}
\author{Yunshan Cao}
\author{Peng Yan}
\email[]{yan@uestc.edu.cn}
\affiliation{School of Physics and State Key Laboratory of Electronic Thin Films and Integrated Devices, University of
Electronic Science and Technology of China, Chengdu 610054, China}

\begin{abstract}
Developed half a century ago, Thiele's equation has played a crucial role in describing the motion of magnetic textures, ranging from simple magnetic domains to exotic magnetic solitons like skyrmions and hopfions. However, it remains a challenge to fully understand the collective dynamics of magnetic texture crystals, due to the complex many-body interactions. On the other hand, electrical circuits have recently been proved to be a powerful platform to realize rich physical phenomena in solids. In this Letter, we first construct a circuit unit to simulate Thiele's equation, which enables us to experimentally observe the ``skyrmion Hall effect". By coupling a pair of such circuit units, we find an emerging frequency comb under strong pumpings, a long-sought nonlinear effect in interacting skyrmion systems. By extending our strategy to circuit arrays, we experimentally observe the topological edge state, thus confirming the theoretical prediction of topological solitonic insulators years ago. Our work builds a faithful connection between electric circuits and magnetic solitons, two seemingly unrelated areas, and opens the door for exploring collective magnetization dynamics in circuit networks.
\end{abstract}
\maketitle

\emph{\color{blue}Introduction}. The Thiele equation was originally proposed by A. Thiele in 1973 to describe the motion of magnetic domains \cite{Thiele1973}, based on the collective-coordinate method for magnetization vector. Over time, it has been recognized that this equation is also powerful in understanding the dynamics of magnetic solitons, like magnetic bubbles \cite{Moon2014,Makhfudz2012}, vortices \cite{Shibata2005}, skyrmions \cite{Fert2017,Yu2021,SChen2021,Tokura2021}, hopfions \cite{Wang2019}, etc. Although Thiele's equation can be exactly solved in the linear regime, its analytical solution in nonlinear cases presents a major challenge, especially when complex interactions among multiple magnetic solitons are involved. Recently, Thiele's equation has been applied to study topological phases in magnetic soliton arrays \cite{Kim2017,Li2018,Li2019,Li2020,Li20202,Li2021,LiPR}. However, due to the difficulty in experimental fabrication and detection, these topological states have not yet been observed \cite{Chen2021}. One may naturally ask: Is there an accessible platform where the intrinsic dynamics is also governed by the Thiele equation such that we are able to experimentally measure these topological phases?

On the other hand, frequency comb (FC) is a spectrum consisting of a series of discrete and evenly spaced spectral lines. The study of FC spans various subjects, including optical \cite{Udem2002,DelHaye2007,Kippenberg2011}, acoustic \cite{Ganesan2017}, mechanical \cite{Jong2023}, and magnonic systems \cite{Wang2021,Wang2022,Xu2023,Shen2024,Liang2024,CWang2024}, induced by typical nonlinear processes, like Kerr nonlinearity in optics and three- and/or four-magnon processes in magnonics. FCs are crucial for exploring fundamental physics and have practical applications in sensing and metrology \cite{Fortier2019}. Recent theories revealed that the FC can appear in interacting skyrmions \cite{Shen2024,Liu2024}, but it is yet to be found experimentally.

Very recently, electrical circuits have become a powerful platform for exploring novel physical phenomena \cite{Yang2024}, such as the simulations of topological states \cite{Imhof2018,Hofmann2019,YWang2020,Song2022}, non-Hermitian dynamics \cite{Helbig2020,Choi2018,Pchen2018}, and non-Abelian gauge fields \cite{NEWu2022}. In this Letter, we first establish a mapping from Thiele's equation to an equivalent circuit response. By constructing a two-node circuit unit consisting of resistors ($R$), capacitors ($C$), and non-reciprocal negative impedance converter (INIC) elements, we showcase the skyrmion Hall effect experimentally, and visualize the inertial effect when the current is switched on, due to the skyrmion mass. The skyrmion velocities and displacements are characterized by node voltages and magnetic
flux, respectively. We then couple a pair of such two-node circuit units by extra inductors, to mimic the dynamics of two interacting skyrmions confined in two adjacent nanodisks. We observe an exotic FC phenomenon under a strong pumping and attribute it to the nonlinearity of the operational amplifiers (OP). Finally, we design a chain of two-node circuit units with alternative connecting inductors to simulate the Su-Schrieffer-Heeger (SSH) array of skyrmions. We observe robust edge modes, confirming the theory of topological soliton insulators. Our work paves the way for simulating topological magnetization dynamics with complex many-body interactions in circuits.

\begin{figure}[t]
  \centering
  \includegraphics[width=0.48\textwidth]{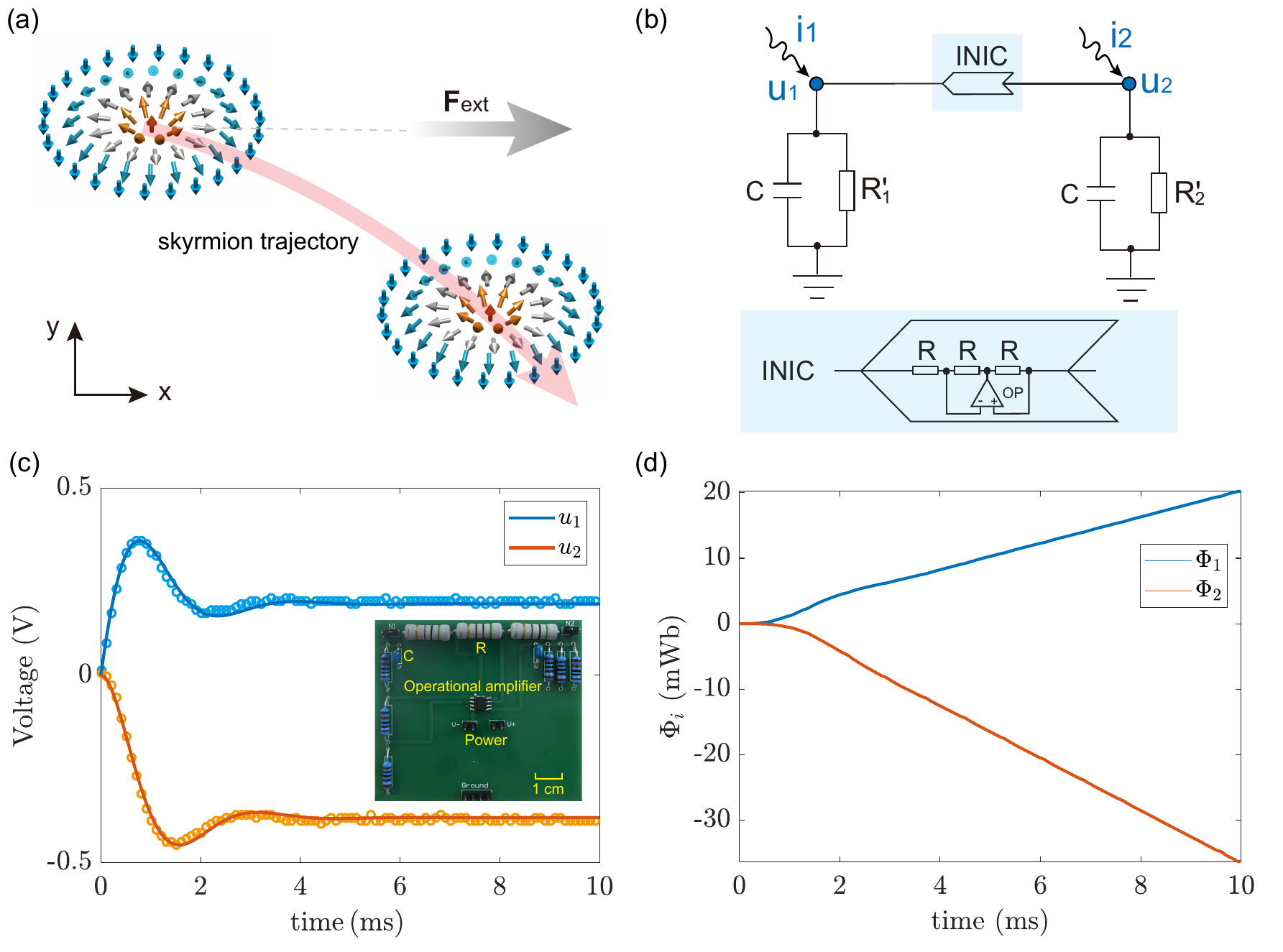}
  \caption{(a) Illustration of the skyrmion Hall effect. (b) Circuit unit to simulate Thiele's equation. The circuit consists of two nodes, coupled by an INIC and grounded by $RC$ elements. INICs are composed of three resistors and an OP, acting as a positive (negative) conductance from left to right (right to left). (c) Voltage responses under a driving on node 1. Circles represent experimental data and curves are theoretical results. Blue and orange symbols correspond to the voltages $u_1$ and $u_2$, respectively. Inset: experimental PCB. (d) Magnetic flux integrated from the voltage responses.}\label{F1}
\end{figure}

\emph{\color{blue} Mapping the Thiele equation to circuit response}. In a rather generic form, the Thiele equation reads
\begin{equation}\label{TE}
-\mathcal{M}\frac{d^2{\bf X}}{dt^2}+\mathcal{G}\hat{z}\times \frac{d{\bf X}}{dt}-\mathcal{D}\alpha \frac{d{\bf X}}{dt}+{\bf F_{\rm ext}}=0,
\end{equation}
where $\mathcal{M}$ is the mass of the magnetic soliton, ${\mathcal G}=4\pi {\mathcal Q}$ represents the gyrotropic coefficient with ${\mathcal Q}$ being the topological charge, $\mathcal{D}$ represents the dissipative coefficient, and $\alpha$ is the Gilbert damping constant. ${\bf X}=(x,y)^{\rm T}$ is the displacement of magnetic solitons and ${\bf F_{\rm ext}}=(F_x,F_y)^{\rm T}$ is the driving force. Without loss of generality, we take the skyrmion as the example in the following circuit design. When the skyrmion is driven by a force along $\hat{x}$ direction, it moves toward an oblique direction, thus exhibiting the skyrmion Hall effect due to its nonzero topological charge, as shown in Fig. \ref{F1}(a).

We then construct an electrical circuit unit to map Thiele's equation above into the circuit response. First of all, we re-write Eq. \eqref{TE} in a matrix form as
\begin{equation}\label{toTE}
{\bf F}_{\rm ext}
=\mathcal{M}\ddot{\bf X}+\mathcal{G}\xi_2\dot{\bf X}+\mathcal{D}\alpha\dot{\bf X},
\end{equation}
where $\dot{\square}$ represents the time derivative and $\xi_2=-i\sigma_2$ with $ \sigma_2$ the Pauli matrix. To find its counterpart in circuit, we consider a two-node circuit shown in Fig. \ref{F1}(b) with the map
\begin{equation}
{\bf i}\leftrightarrow{\bf F}_{\rm ext}~{\rm and}~{\bf u}\leftrightarrow\dot{\bf X},
\end{equation}
where ${\bf i}=(i_1,i_2)^{\rm T}$ and ${\bf u}=(u_1,u_2)^{\rm T}$ represents the current flowing into the $n$th node ($n=1,2$) and voltage at the $n$th node, respectively. An INIC unit is used to simulate the gyrotropic coefficient $\mathcal{G}$. The resistors $R_1$ and $R_2$ represent the dissipative term $\mathcal{D}\alpha$ and the capacitor $C$ mimics the skyrmion mass $\mathcal{M}$. The governing equation for such a circuit can be derived using the nodal-analysis method, as follows
\begin{equation}\label{Eq3}
{\bf i}=C\dot{\bf u}+g\xi_2{\bf u}+g'{\bf u},
\end{equation}
with $g=1/R$ and $g'=(-1)^{n+1}1/R+1/R'_n$ being the conductances. Experimentally, we choose $g=2$ mS ($R_2=0.5$ k$\Omega$), $g'=1$ mS ($R'_1=-1$ k$\Omega$ realized by an INIC unit and $R'_2=333$ $\Omega$), and $C=1$ $\mu$F.

To validate the mapping, we apply a driving current on node 1 in the form $i_1(t)=i_0H(t)$ with $i_0=951$ $\mu$A and $H(t)$ being a Heaviside step function. The steady-state voltage can be calculated by the formulas $u_1
=\frac{g'}{g'^2+g^2}i_0=0.19$ V and $u_2=\frac{-g}{g'^2+g^2}i_0=-0.38$ V. This design is analogous to imposing a force along $\hat{x}$-direction on the skyrmion. The skyrmion then moves along both $\hat{x}$ and $-\hat{y}$ direction (manifesting as the skyrmion Hall effect with topological charge $\mathcal{Q}=1$). An oscillation emerges in the initial stage due to the finite mass carried by the skyrmion. A print circuit board (PCB) is built to measure the voltage signal, as plotted in Fig. \ref{F1}(c). It clearly shows the time-dependence of voltage signals at both nodes, with initial oscillations caused by the capacitors, and the steady-state values in the long run. Experimental results (symbols) compare well with the theoretical calculations (curves). 

Meanwhile, it noted that the skyrmion displacement is represented by the negative magnetic flux, defined as $\Phi_n=\int_{-\infty}^t dt' u_n(t')$ (Wb). Figure \ref{F1}(d) displays the time evolution of the magnetic flux at each node, from which one observes a positive ``displacement" $\Phi_1$ at node 1 and a negative Hall ``displacement" at node 2 with the Hall angle $\theta_{\rm Hall}=\arctan\left(\frac{u_2}{u_1}\right)=-63^\circ$. These experimental results demonstrate that the circuit model fully captures the essential physics governed by Thiele's equation.

\begin{figure} 
  \centering
  \includegraphics[width=0.48\textwidth]{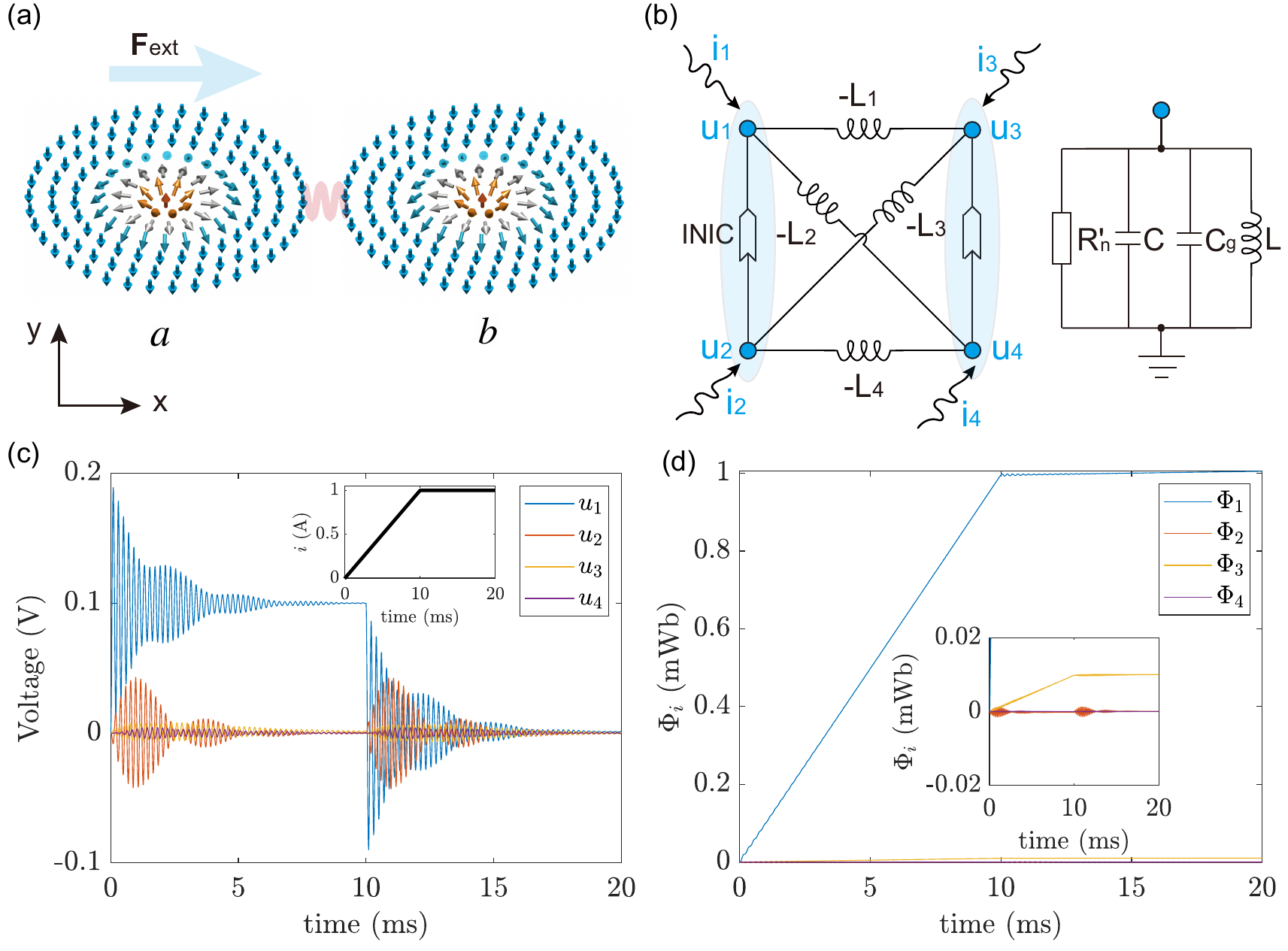}
  \caption{(a) Illustration of coupled skyrmions $a$ and $b$. (b) Circuit model with node 1, 2 and 3, 4 simulating the motion of skyrmion $a$ and $b$, respectively. (c) Voltage responses  under the driving current on node 1. Inset: time-dependence of the current density. (d) Magnetic flux by integrating the voltage. Inset: zoom in the magnetic fluxs near zero.}\label{F2}
\end{figure}

\emph{\color{blue} Skyrmion-skyrmion interaction.}
Next, we incorporate the skyrmion-skyrmion interaction and the boundary potential into the circuit model. Considering two coupled skyrmions $a$ and $b$ in the nanodisks as shown in Fig. \ref{F2}(a), the energy of this system is given by 
$E=\frac{K}{2}({\bf X}_a^2+{\bf X}_b^2)+ E_{ab}$ \cite{Kim2017}.
${\bf X}_j=(x_{j},y_{j})^{\rm T}$ ($j=a,b$) is the displacement of the $j$-th skyrmion from its equilibrium position. The first term parametrizes the boundary potential by the spring constant $K>0$. The second term is the interaction between two skyrmions $E_{ab}=I_{\|}X^{\|}_aX^{\|}_b-I_\perp X_a^{\perp}X_b^{\perp}$
with $I_{\|}$, $I_\perp$ parametrizing the interaction. Here, $X^{\|}_{j}=\hat{e}_{ab}\cdot{\bf X}_{j}$ and $X^\perp_{j}=(\hat{z}\times\hat{e}_{ab})\cdot{\bf X}_{j}$ with $\hat{e}_{ab}$ the unit vector from skyrmion $a$ to $b$. Defining $\theta$ as the angle of the direction $\hat{e}_{ab}$ from the $\hat{x}$ axis, 
the force ${\bf F}_{a,b}=-\frac{\partial E}{\partial {\bf X}_{a,b}}$ is expressed as
$\left(
  \begin{array}{c}
    F_{ax(bx)} \\
    F_{ax(by)} \\
  \end{array}
\right)=-K\left(
  \begin{array}{c}
    x_{a(b)} \\
    y_{a(b)} \\
  \end{array}
\right)-\mathscr{I}\left(
  \begin{array}{c}
    x_{b(a)} \\
    y_{b(a)} \\
  \end{array}
\right),$
where $\mathscr{I}=\left(
                \begin{array}{cc}
                  \zeta+\xi\cos(2\theta) & \xi\sin(2 \theta)\\
                  \xi\sin(2 \theta) & \zeta-\xi\cos(2\theta)\\
                \end{array}
              \right)$ is the interaction matrix with $\xi=I_{\|}+I_\perp$ and $\zeta=I_{\|}-I_\perp$.

One can write the Thiele equation of the two interacting skyrmion as 
\begin{equation}\label{Eq5}
{\bf F}_{\rm ext}=\mathcal{M}\ddot{\bf X}+\mathcal{G}\xi_4\dot{\bf X}+\mathcal{D}\alpha\dot{\bf X}-{\bf F}_{\rm int}
\end{equation}
where ${\bf F}_{\rm int}=P{\bf X}$  with $P=-\left( \begin{array}{cc}
  K\sigma_0 & \mathscr{I} \\
   \mathscr{I} & K\sigma_0 \\
    \end{array}
\right)$ and ${\bf X}=(x_a,x_b,y_a,y_b)^{\rm T}$ arises from both the boundary potential and skyrmion-skyrmion interaction and ${\bf F}_{\rm ext}$ is the external driving force.

With the substitution ${\bf \Phi}=(\Phi_1,\Phi_2,\Phi_3,\Phi_4)^{\rm T}\rightarrow {\bf X}$, ${\bf u}=(u_1,u_2,u_3,u_4)^{\rm T}\rightarrow \dot{{\bf X}}$, and ${\bf i}=(i_1,i_2,i_3,i_4)^{\rm T}\rightarrow {\bf F}_{\rm ext}$, the circuit Thiele equation is expressed as
\begin{equation}\label{TtwoS}
{\bf i}=C\dot{\bf u}+g\xi_4{\bf u}+g'{\bf u}+L^{-1}P_L{\bf \Phi},
\end{equation}
where $\xi_4=\sigma_0\otimes\xi_2$ and $P_L=\left[\sigma_0\otimes\sigma_0+\sigma_1\otimes\left(\begin{array}{cc}
 r_1 & r_2\\
 r_3 & r_4\\
\end{array}\right)\right]$
with $r_n=L/L_n$ ($n=1,2,3,4$). Inductors $L$ and $L_n$ are adopted to simulate the force from boundaries and interactions, respectively.

We therefore can construct the circuit model by writing Eq. \eqref{TtwoS} in frequency space, and the corresponding circuit diagram is illustrated in Fig. \ref{F2}(b) with four negative inductors $L_n$ describing the interactions. The negative sign arises because the mutual conductance for an inductor is $-\frac{1}{i\omega L_n}$. We choose the spring constant $K$ as $1/L=1$ mH$^{-1}$ and suppose $\mathscr{I}=\left(
                       \begin{array}{cc}
                         -0.01 & 0 \\
                         0 & 0.1 \\
                       \end{array}
                     \right)K
$, and these parameters are close to the results reported in Ref. \cite{Li2018}. In circuits, we set $L_1=-100$ mH, $L_2=L_3=0$, and $L_4=10$ mH. Analogy to the gyration frequency of skyrmion $\omega_0=K/|G|$, we set the circuit eigen frequency as $f_0=\omega_0/(2\pi)=1/(2\pi gL)=79.6$ kHz. It's noted that $L_4=10$ mH is positive but we need a negative inductor, realized by a capacitor $C_4=\frac{1}{\omega_0^2L_4}=0.4$ nF.

Next, we establish the circuit model and measure the voltage response by adding a driving on node 1, with the current evolution plotted in the inset of Fig. \ref{F2}(c). For $t<10$ ms, as the current density increases, one can see that the voltage rises rapidly but becomes oscillating near $u_1=L\frac{di}{dt}=0.1$ V, due to both the limit of inductor $L$ (acting as the boundary potential) and the mass effect. For $t>10$ ms, the voltage drops to zero because the circuit is grounded by inductor directly for a constant current (the driving force on skyrmion is compensated by the repulsion force from the boundary). The finial position of skyrmions can be estimated by Hooke's law with ${\bf F}_a={\bf F}_{\rm ext}$ and ${\bf F}_b={\bf F}_{\rm int}$, which gives $\Phi_1=I_1/K=1 {\rm A}/(1~{\rm mH}^{-1})=1~{\rm mWb}$ and $\Phi_3=0.01~{\rm mWb}$.

Figure \ref{F2}(d) shows the magnetic flux (skyrmion position) as a function of time, and a $\hat{x}$-direction displacement $\Phi_1$ of skyrmion $a$ is found accompanied by a $\hat{x}$-direction displacement $\Phi_3$ of skyrmion $b$ caused by the inter-skyrmion coupling. The final displacements are consistent with theoretical estimations above.

\begin{figure}
  \centering
  \includegraphics[width=0.48\textwidth]{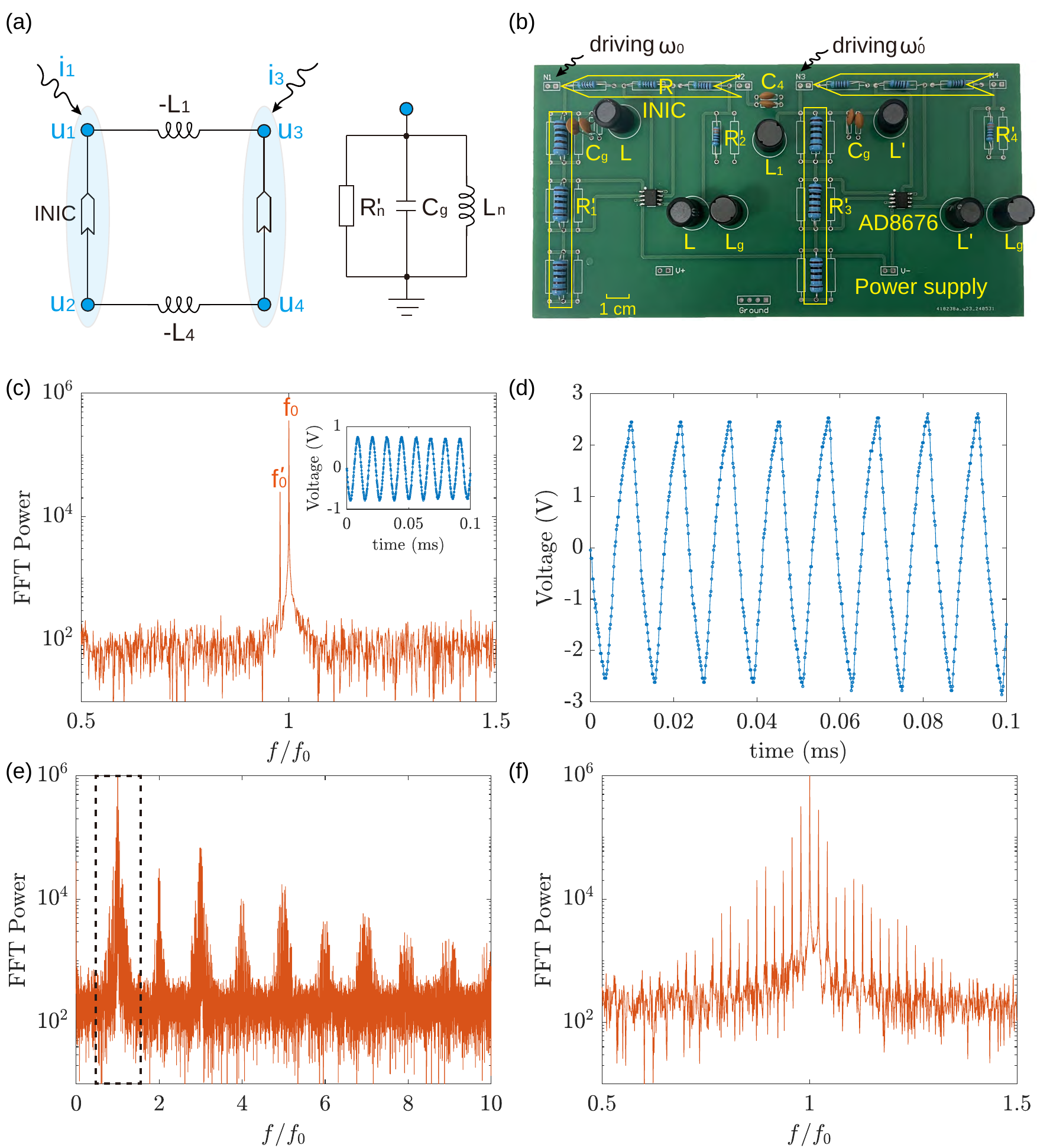}
  \caption{(a) Illustration of circuit model for two coupled skyrmion nanodisks. (b) Experimental PCB. (c) FFT results of the voltage response on node 2 for small signal. Inset: experimental voltage signal (dots are the output from oscilloscope and lines are obtained by interpolation method). (b) Experimental  voltage response on node 2 for $V_0=3$ V. (e) FFT results for the signal in (d). (f) FC by zooming in the frequency range $f\in[0.5,1.5]f_0$. }\label{F3}
\end{figure}

\emph{\color{blue}Circuit frequency comb.} Then, we consider two coupled skyrimons with slightly different gyration frequencies $\omega_0$ and $\omega'_0$ ($\omega_0\approx\omega_0'$), the circuit model of which is shown in Fig. \ref{F3}(a). Here, we omit the mass of skyrmion, i.e., the first term in the right side of Eq. \eqref{TtwoS}, for simplicity, and the circuit equation reads
\begin{equation} \label{PhiE}
\dot{\bf \Phi}+\chi^{-1}\Omega P_L{\bf \Phi}=g^{-1}\chi^{-1}{\bf i},
\end{equation}
where $\Omega={\rm diag}\{\omega_0,\omega_0,\omega'_0,\omega'_0\}$ and $\chi=\sigma_0\otimes(\frac{g'}{g}\sigma_0-i\sigma_2)$.
Diagonalizing the coefficient matrix $\chi^{-1}\Omega P_L$ of ${\bf \Phi}$ with the matrix transform $D\Lambda D^{-1}$ ($\Lambda$ is a diagonal matrix and $D$ is eigenmodes of the coefficient matrix), we obtain
\begin{equation} \label{PsiE}
\frac{d{\bf \Psi}}{dt}+\Lambda {\bf \Psi}={\bf j},
\end{equation}
with ${\bf \Psi}=D^{-1}{\bf \Phi}$ and ${\bf j}=D^{-1}g^{-1}\chi^{-1}{\bf i}$.
We suppose the two driving currents as $i_1(t)=I_0\sin(\omega_0 t)$ and $i_3(t)=I_0\sin(\omega'_0 t)$. The formal solution of Eq. \eqref{PsiE} is given by
\begin{equation}\label{Psi}
\Psi_n(t)=c_ne^{-\int \Lambda_ndt'}+e^{-\int\Lambda_ndt'}\int j_n(t') e^{\int\Lambda_ndt''} dt',
\end{equation}
where $\Lambda_n$ is the $n$-th diagonal elements of $\Lambda$ and $c_n$ is a coefficient. Performing a fast Fourier transformation (FFT) on the solutions, one should expect merely two peaks at $f_0$ and $f_0'$. 

We print a circuit board to examine this problem, as shown in Fig. \ref{F3}(b). In experiments, we choose $L=938$ $\mu$H, $L'=958$ $\mu$H, $L_1=10$ mH, and $C_4=352$ pF, measured by an Impedance Analyzer. The resonant frequencies are $f_0=1/(2\pi gL)=84.8$ kHz and $f_0'=1/(2\pi gL')=83$ kHz, respectively. Firstly, imposing the driving $u_1=u_0\sin(\omega_0t)$ and $u_3=u_0\sin(\omega_0't)$ with $u_0=1$ V on nodes 1 and 3, respectively, the voltage signal of node 2 is shown in the inset of Fig. \ref{F3}(c), and one can see two frequency peaks at $f_0$ and $f_0'$ in frequency space, as displayed in Fig. \ref{F3}(c), which is consistent with the prediction of linear theory. 

However, the voltage response of node 2 suffers with a deformation under a strong pumping with $u_0=3$ V, as shown in Fig. \ref{F3}(d). The voltage transforms from a beat frequency signal to a triangle-like one, which is attributed to the nonlinear effect of OPs, i.e., the limit of slew rate \cite{Allen1977,Allen1978}, see Supplemental Material (SM) for details \cite{SM}. By imposing FFT on the signal, we obtain the spectrum in Fig. \ref{F3}(e), and a series of frequency multiplication signals emerge due to the circuit nonlinearity. The odd-order signals are stronger than the even order ones, because the FFT for a triangle wave with frequency $\omega$ generate $(2m-1)\omega$ ($m=1,2,3,...$). The even-order signals arise because of the frequency doubling. If one zooms in the frequency spectrum near $mf_0$, a regular FC can be clearly identified [see Fig. \ref{F3}(f)], and the interval is determined by the difference between gyration frequencies of two skyrmions. 

\begin{figure}
  \centering
  \includegraphics[width=0.48\textwidth]{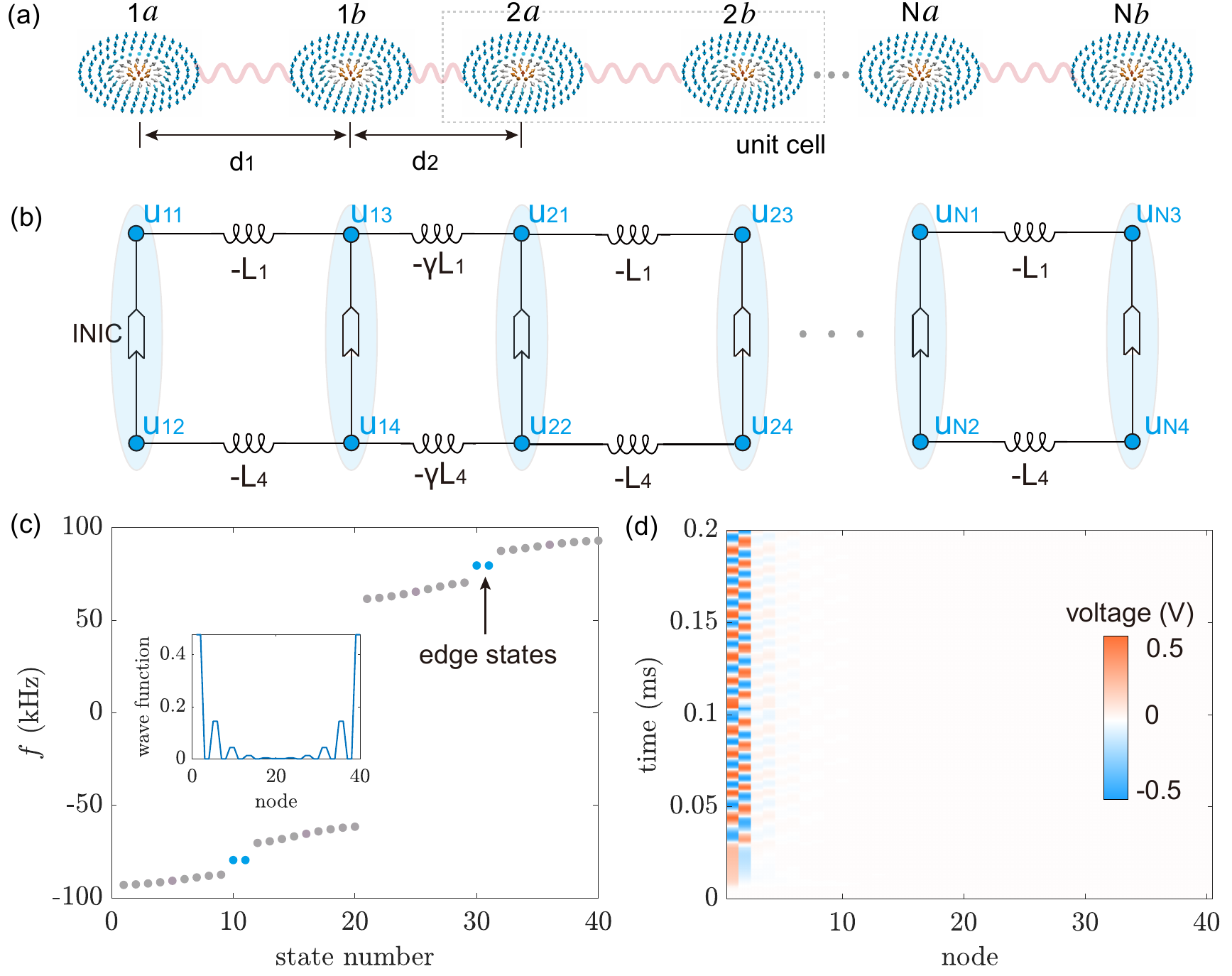}
  \caption{(a) Illustration of a skyrmion SSH model. Dashed rectangle indicates the unit cell with two skyrmions $a$ and $b$. (b) Circuit chain with coupled two-node circuits. Each circuit node is grounded by the same $RLC$ element shown in Fig. \ref{F3}(a). (c) Eigenvalues for the circuit. The gray and blue dots represent the bulk and edge states, respectively. Inset: wave functions for the edge state. (d) Time evolution of the edge state.}\label{F4}
\end{figure}

\emph{\color{blue}Circuit realization of skyrmion lattices.}
Finally, we demonstrate an extended circuit to verify the theoretical prediction of topological edge states emerging in soliton arrays. Here, we take a $2N$-skyrmion SSH model as an example \cite{Go2020}, as shown in Fig. \ref{F4}(a). This skyrmion dynamics of the $n$th unit cell is described by the Thiele equation set
\begin{equation}
\mathcal{G} \xi_4 \dot{{\bf X}}_n+\mathcal{D}\alpha\dot{{\bf X}}_n=P{\bf X}_n
+P^-{\bf X}_{n-1}+P^+{\bf X}_{n+1}+{\bf F}_{\rm ext},
\end{equation}
where ${\bf X}_n=(x_{na},y_{na},x_{nb},y_{nb})^{\rm T}$ are the displacement of skyrmion in the $n$th unit cell. The matrices $P$ [same as the one in Eq. \eqref{Eq5}] and 
$P^{\pm}=\frac{1}{2\gamma}(\sigma_0\pm\sigma_3)\mathscr{I}$
describe the intracell and intercell skyrmion interactions, respectively.

One can map the skyrmion chain with the circuit in Fig. \ref{F4}(b), governed by the circuit equation
\begin{equation}
g\xi_4\dot{{\bf \Phi}}_n+g'\dot{{\bf \Phi}}_n=-g\omega_0P_L{\bf \Phi}_n
-g\omega_0P_L^-{\bf \Phi}_{n-1}-g\omega_0P_L^+{\bf \Phi}_{n+1}+{\bf i}_n,
\end{equation}
with ${\bf \Phi}_n=(\Phi_{n1},\Phi_{n2},\Phi_{n3},\Phi_{n4})^{\rm T}$ representing the magnetic fluxes of four node in the $n$th circuit unit cell. Here, $P_L$ is equal to the one in Eq. \eqref{TtwoS} and 
$P^{\pm}_L=\frac{1}{2\gamma}(\sigma_0\pm\sigma_3)\left(
                                                        \begin{array}{cc}
                                                          r_1 & 0 \\
                                                          0 & r_4 \\
                                                        \end{array}
                                                      \right)$.
We compute the band structures and evaluate the Zak phase for an infinitely long chain, and find that this system allows topologically nontrivial phase for $\gamma<1$ with $\gamma$ the hopping ratio \cite{SM}.

We then calculate the eigenvalues of a finite circuit with $N=10$ unit cells and plot the frequency spectrum in Fig. \ref{F4}(c) for $\gamma=1/3$. Two isolated modes (blue dots) are identified inside the band gap, corresponding to the edge states of the SSH soliton model. Next, we perform a time-resolved measurement on the edge state by adding a voltage signal $u_1=0.5\sin(\omega_0 t)$ (V) at the node 1. One can observe a stable edge state at the sample edge, as shown in Fig. \ref{F4}(d). We have checked that the edge mode is quite robust against imperfections of circuit elements. This extended circuit successfully simulates the skyrmion array with non-trivial edge modes.

\emph{\color{blue}Discussion and conclusion.}
With the modified Thiele equation, one can explore other rich features of magnetic solitons. For instance, one can investigate the non-Newtonian motion with extra circuit elements \cite{Ivanov2010}, and the finite-temperature effect by introducing element fluctuations in circuits \cite{Zhao2020,Wei2021}. The circuit networks can also enable us to study the dislocation effect in imperfect skyrmion crystals \cite{Okuyama2024}. Besides, one can extend the circuit model to higher-dimensional cases to verify the theoretical predictions of skyrmion based hinge states \cite{Li20212} or surface arc \cite{Li2022}. At present, the interplay between the nonlinearity and topology has aroused many attentions \cite{Ricketts2018,Hadad2017,Hadad2018,Wang1102,Tao2020,Hohmann2023,Tang2023}. Its manifestation in skyrmion lattice is also an interesting topic from both the theoretical perspective and experimental examination in circuit platform.

To conclude, we mapped the classic Thiele equation to circuit responses. By constructing a two-node circuit unit, we experimentally demonstrated the intriguing skyrmion Hall effect in circuits. We then designed circuits to simulate two interacting skrymions. Interestingly, for such a circuit-unit with different driving frequencies, we observed an emerging circuit FC under a strong pumping, a long-sought phenomenon in interacting-skyrmion systems. We finally verified the theory of topological soliton insulators in an array of circuit units with alternating inductor connections by experimentally observing the robust edge states. Our work builds a fruitful bridge between topological magnetization dynamics and electric circuits, which also opens the door for exploring exotic many-body physics in magnetism based on circuit platform.

\begin{acknowledgments}
We thank Zhejunyu Jin for helpful discussion. This work was supported by the National Key R\&D Program under Contract No. 2022YFA1402802 and the National Natural Science Foundation of China (Grants No. 12374103, No. 12434003, and No. 12074057). Lingling Song acknowledges the financial support of the China Postdoctoral Science Foundation (Grant No. 2024T170095).
\end{acknowledgments}

\end{document}